\documentclass[preprint,superscriptaddress,
frontmatterverbose,showpacs,preprintnumbers,amsmath,amssymb,aps,showkeys]{revtex4-1}
\usepackage{graphicx} 
\usepackage{bm} 
\usepackage[pagebackref=false,colorlinks,linkcolor=red,citecolor=magenta]{hyperref} 
\begin{document} 
\title{A theoretical scheme for the realization of the sphere-coherent motional states in an atom-assisted optomechanical cavity} 
\author{F. Bemani} \email{foroudbemani@gmail.com} 
\affiliation{Department of Physics, Faculty of Science, University of Isfahan, Hezar-Jerib st.,  Isfahan, Iran} 
\author{R. Roknizadeh} \email{rokni@sci.ui.ac.ir} 
\author{M. H. Naderi} \email{mhnaderi@phys.ui.ac.ir} 
\affiliation{Department of Physics, Quantum Optics Group, University of Isfahan, Hezar-Jerib st.,  Isfahan, Iran} 
\date{\today} 
\begin{abstract} 
A theoretical scheme for the realization of the sphere-coherent motional states in an optomechanical cavity in the presence of a two-level atom is proposed. To this end, the analogy between an atom-assisted optomechanical cavity and a laser-driven trapped-ion system is used. This analogy provides us with a theoretical tool to show how sphere-coherent states can be generated for the motional degree of freedom of the macroscopic mechanical oscillator from atom-field-mirror interactions in a multi-mode optomechanical cavity. Some nonclassical properties of the generated state of the mechanical oscillator, including the degree of quadrature squeezing and the negativity of the Wigner distribution are studied. We also examine the effects of the dissipation mechanisms involved in the system under consideration, including the atomic spontaneous emission and the damping of the motion of the mechanical oscillator, on the generated motional sphere-coherent states. 
\end{abstract} 
\maketitle 
\section{\label{sec:SecI}Introduction} 
The coupling of photons and phonons via radiation pressure in the optomechanical systems (OMSs) leads to the entanglement and momentum exchange between electromagnetic radiation and macroscopic (or mesoscopic) mechanical oscillators (MOs). The cavity optomechanics has been investigated intensively from both theoretical and experimental point of views in the last few years (for an extensive recent review, see \cite{Marq}). There are several motivations to study this emerging field of research. OMSs enable us to analyze the quantum behavior of macroscopic systems and their coherent interaction with light. In addition, cooling of the mechanical motion to the quantum ground state \cite{Teufel,Chan}, ultra-sensitive detection \cite{Anetsberger, Verlot, Teufel2}, entanglement \cite{Ghobadi} and generation of non-classical states for light and MO \cite{Brooks} are major themes that have received much attention. Using the radiation pressure force, one can manipulate the MO motion and bring it into the quantum regime. The generation of Fock states \cite{OConnell} and their superpositions \cite{Xu} for the macroscopic MO is of great importance. Theoretical schemes for producing Schr\"{o}dinger cat states as well as their nonlinear multi-component counterparts, and other types of nonclassical states of the MO have also been proposed \cite{Bose,Barzanjeh, Zheng}. As the realization of nonclassical states of MO is important for the foundation of physics, OMSs play an important role in the exploring of the boundaries between the classical physics and the quantum mechanics. 
 
The hybrid OMSs, with single or ensemble of atoms, coupled to the MO in the quantum regime have recently become the subject of an intense research effort \cite{ Treutlein,Hammerer,Wallquist,Hunger,Ian}. Using the atomic physics toolbox can upgrade and extend the field of Optomechanics. The simplest type of hybridization is the presence of a single two-level atom in the optomechanical cavity. Both internal and motional degrees of freedom of the atom can be coupled to the motion of the MO. The presence of an atom in the optomechanical cavity can increase the strength of the coupling between the optical and the mechanical degrees of freedom. In this manner, 
the quantum behavior in an OMS can be observed and controlled more efficiently. 
Recently, the analogy between a single atom in the optomechanical cavity and a trapped ion has been shown by using the Susskind-Glogower phase operator \cite{moyacessa}. Moreover, it has been shown \cite{Xu} that a driven bare OMS in which the classical driving field couples only two lowest energy levels of the cavity field, is analogous to a system of a single two-level trapped cold atom interacting with a classical driving field and vibrating along one direction. These analogies provide a convenient theoretical tool to investigate OMSs by methods and techniques that have been used for exploration of the trapped ion systems.  
 
One of the most important features of the OMSs is a kind of intrinsic nonlinearity resulting from the mutual interaction between the cavity field and the MO \cite{Gong}. This nonlinearity can play a key role in the generation of nonclassical states of both the cavity field and the MO. In connection with nonclassical states, much attention has recently been paid to the deformed (nonlinear) coherent states because of their relevance in nonlinear quantum optics. From the mathematical point of view, they correspond to nonlinear algebras rather than Lie algebra \cite{Manko}. The physically significant feature of these states and their superpositions is their nonclassical characteristics such as amplitude squeezing and sub-Poissonian statistics\cite{Sivakumar}. The nonlinear coherent states are not merely mathematical objects. By proposing a scheme based on the strong nonlinearities inherent in the Jaynes-Cummings Hamiltonian for the laser-assisted vibronic interaction \cite{Vogel}, it has been shown \cite{de Matos} that these states may appear as stationary states of the center-of-mass motion of laser-driven trapped ion far from the Lamb-Dicke regime. In Ref. \cite{mahdifar2}, by using the
nonlinear coherent states approach, the algebra of a two-dimensional harmonic oscillator on the flat surface as well as on a sphere has been studied \cite{mahdifar2}. In this way, the authors have introduced a special
type of these states, the so-called sphere-coherent states (SCSs), and studied their nonclassical properties. Recently, it has been proposed \cite{mahdifar} a physical scheme that allows one to generate and control the SCSs, as the motional dark states of a properly laser-driven trapped-ion system. As a key result, the authors have shown that by adjusting the Rabi frequencies and the phases of the lasers driving the ion, one can simulate and control the curvature of the space in which the SCSs are prepared.  
Motivated by the great possibilities the OMSs are expected to produce nonclassical states of both the MO and the cavity field as well as the analogy between an atom-assisted OMS and a laser-driven
trapped-ion system, in the present contribution we deal with the question of how the SCSs can be generated in a multimode optomechanical cavity. The nonlinear character of the OMS plays an important role in our treatment. The system under consideration is a hybrid system formed by a single
two-level atom which is trapped in a multimode Fabry-Perot cavity with a vibrating end mirror. We assume that the internal states of the atom are
indirectly coupled to the MO vibrations via the common interaction with the cavity field. The system is affected by various dissipation mechanisms. Separating time scales of dissipations allows us to generate the SCSs for time scales much larger than the atomic characteristic time and smaller than the MO characteristic time of decay. The $N$-dimensional SCSs can be achieved using several optical modes properly coupled to the internal states of the single atom and the motional degree of freedom of the MO. These states are highly nonclassical, i.e., they show quadrature squeezing and they have Wigner function with negative values in some regions. The curvature of the sphere is determined by the intensities of the optical fields as well as the coupling strength of atom-fields, which can be controlled experimentally. Furthermore, the dimension of the Hilbert space can be controlled by the number of the optical modes in the cavity. The influence of the damping of the motion of the MO on the generation of the motional SCSs is also studied. 

The paper is organized as follows. In Sec. \ref{sec:SecII}, we derive an effective Hamiltonian for the optomechanical system under consideration. In Sec. \ref{sec:SecIII} we make a brief review on the SCSs and their nonclassical features. In Sec. \ref{sec:SecIV}, we present and discuss our proposal for the realization of the SCSs in the single-atom assisted OMS. 
In Sec. \ref{sec:SecV}, we examine the effect of the damping of the motion of MO on the generated SCSs. Finally, in Sec. \ref{sec:SecVI} we summarize our findings and conclude with a future outlook. 
\section{\label{sec:SecII} System Hamiltonian} 
\begin{figure*}
\centering 
\includegraphics[width=14cm]{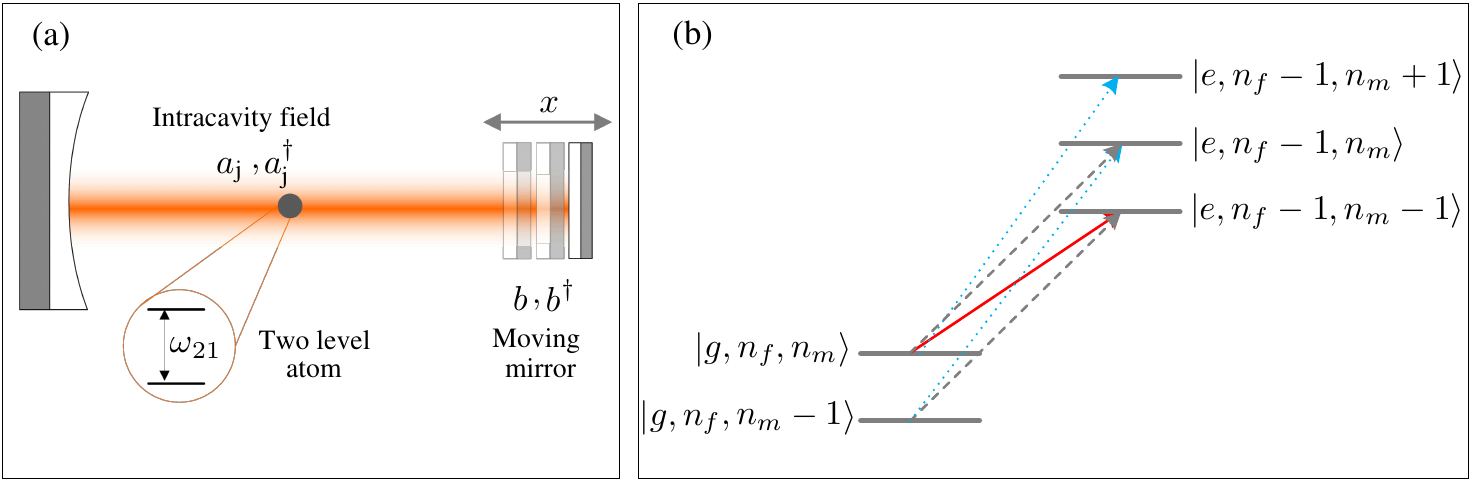} 
\caption{(Color online) (a) A Fabry-Perot cavity with a vibrating end mirror with frequency $\nu$, as the mechanical element, containing a two-level atom with transition frequency $\omega _{21}$. The MO vibrations are coupled to the atomic internal states by the cavity field. (b) Level diagram structure for an OMS in the presence of a single two-level atom. Possible transitions correspond to the Hamiltonian (\ref{InteractionHamiltonian2}) with the condition ${\Delta _j}=\pm q \nu $ (see the text). The dashed, solid, and dotted arrows display carrier transition $(q=0)$, the first red $(q=1)$ and the first blue $(q=-1)$ sidebands, respectively.} 
\label{fig:FIG1} 
\end{figure*}
As depicted in Fig. (\ref{fig:FIG1}-a), we consider a hybrid optomechanical system formed by a single two-level atom with transition frequency $\omega_{21}$ which is trapped in a Fabry-Perot cavity with a movable end mirror vibrating with frequency $\nu$. The motion of the MO is coupled to the optical field via radiation pressure force. The coupling between the internal degree of freedom of the atom and the cavity field is described by the Jaynes-Cummings interaction. In general, the atom-field coupling constant depends on the position of the atom, however, by applying the electric dipole approximation the spatial dependence can be neglected. The coupling of the atom and the mechanical oscillator is mediated by the cavity field. Here, we restrict our consideration to the case of a single-mechanical and cavity modes. For the cavity field, the single-mode assumption is valid if the cavity free spectral range is much larger than the
mechanical frequency $\nu$ \cite{Law}. Restriction to a single mechanical mode is justified when the detection bandwidth is chosen such that it includes only a single, isolated,
mechanical resonance and mode-mode coupling is negligible \cite{Genes}. Later, in Sec. \ref{sec:SecIV}, we generalize our consideration to the case of multimode cavity field to investigate the possibility of the generation of the motional SCSs.  
 
The total Hamiltonian of the system is given by ($\hbar=1$) 
\begin{eqnarray} 
&&\mathcal{H}_{1}= \nu {{ b}^\dag } b + 
\frac{1}{2}{\omega _{21}}{ \sigma ^z} + {\omega _j}{{n}_j} \nonumber\\ 
&&\qquad \qquad + {h _j} ( {{{ a}_j} + a_j^\dag ) ( { \sigma ^ + }+{ \sigma ^ - }})- {g_j}{{ n}_j}( { b + {{b}^\dag }} ),\qquad 
\label{eq:Hamiltonian1} 
\end{eqnarray} 
where the mechanical (optical) mode of frequency $\nu$ ($\omega_j$) is described by the usual bosonic annihilation and creation operators $b$, $b^\dag$ ($a_j$, $a_j^\dag$). Here, the two-level atom with transition frequency $\omega _{21}$ is described by the spin-$1/2$ algebra of the Pauli matrices $\sigma ^ +$, $\sigma ^ -$ and $\sigma ^ z$. The atom-cavity coupling strength and optomechanical coupling constant are given by $h_j$ and $g_j$, respectively. 

It is now convenient to rotate the Hamiltonian (\ref{eq:Hamiltonian1}) using the polaron transformation \cite{Polaron1} given by  
\begin{equation} 
\mathcal{D}_j \left( {\alpha_j ,{{n}_j}} \right) = {e^{\alpha_j {{n}_j}\left( {{{ b}^\dag } -  b} \right)}}, 
\end{equation} 
in which $\alpha _j$ is the effective Lamb-Dicke parameter, defined as the ratio of the optomechanical coupling, $g_j$, and the mirror vibrational frequency, $\nu$. The rotated Hamiltonian, ${{\cal H}_2} = {{\cal {D}}^\dag }{{\cal H}_1}{\cal D}$, reads 
\begin{eqnarray} 
{\cal H}_2={\cal H}_2^{f} + {\cal H}_2^{i},
\label{totHamiltonian} 
\end{eqnarray} 
where ${\cal H}_2^{f}$ and ${\cal H}_2^{i}$ are, respectively, the free and interaction parts of the Hamiltonian: 
\begin{equation} 
{\cal H}_2^f = \nu {{ b}^\dag } b + \frac{1}{2}{\omega _{21}}{ \sigma ^z} - \frac{{g_j^2}}{\nu } n_j^2 + {\omega _j}{{ n}_j}, 
\label{fHamiltonian} 
\end{equation} 
\begin{equation} 
{\cal H}_2^i = {h_j}\left[ {{{\cal D}_j}\left( \alpha _j,{1} \right){{a}_j}{ \sigma ^ + } +  a_j^\dag { \sigma ^ - }{\cal {D}}_j^\dag \left( \alpha _j,{1} \right)} \right]. 
\label{iHamiltonian} 
\end{equation} 
The nonlinear characteristic of the transformed Hamiltonian, ${\cal H}_2$ is evident. In fact, the radiation pressure introduces an effective Kerr nonlinearity in the Hamiltonian (\ref{fHamiltonian}) which make it possible to define an intensity-dependent detuning \cite{Kerr}. Furthermore, the atom-field coupling appears as an intensity-dependent one which is dependent also on the MO variables. The Hamiltonian (\ref{totHamiltonian}), is the starting point of our discussion in this paper. 
The polaron (displacement) operator can be expanded as follows \cite{Haroche}: 
\begin{eqnarray} 
&&{{\cal D}_j}\left( {{\alpha _j}},{1} \right)= 
 \sum\limits_{q > 0} {\left[ {{f_q}\left( {{{n}_m},{\alpha _j}} \right){{\left( { - {\alpha _j}} \right)}^q}{{b}^q} + {\alpha ^q}{{b}^\dag }^q{f_q}\left( {{n_m},{\alpha _j}} \right)} \right]} \nonumber \\ 
&& \qquad \qquad \qquad \qquad + {f_0}\left( {{{n}_m},{\alpha _j}} \right), 
\end{eqnarray} 
where the function ${f_q}\left( {{n_m},{\alpha _j}} \right)$ has the form 
\begin{equation} 
{f_q}\left( {{n_m},{\alpha _j}} \right) = {e^{ - \frac{1}{2}|{\alpha _j}{|^2}}}\sum\limits_{l = 0} {\frac{{\alpha _j^{2l}{{\left( { - 1} \right)}^l}}}{{l!(l + q)!}}:{n^l}_m:}, 
\end{equation} 
and $: :$ notation denotes normal ordering. In the following, we consider the Hamiltonian ${\cal {H}}_2^i$ in the interaction picture, ${{ {\cal {\tilde{H}}}}_2} = {e^{ i{\cal {H}}_2^ft}}{\cal {H}}_2^i{e^{-i{\cal {H}}_2^ft}}$, that reads 
\begin{widetext} 
\begin{eqnarray} 
&&{{ {\cal {\tilde{H}}}}_2} = {h_j}\sum\limits_{q > 0} {{{\left( { - {\alpha _j}} \right)}^q}{{ a}_j}\left( {{  \sigma ^ + }{e^{i\left( {{\Delta _j} - q\nu } \right)t}} + { \sigma ^ - }{e^{i\left( {{\Delta _j} - q\nu - 2{\omega _{21}}} \right)t}}} \right){f_q}\left( {{{ n}_m},{\alpha _j}} \right){{ b}^q}} \nonumber \\ 
&&\qquad + {h_j}\sum\limits_{q > 0} {{\alpha ^q}{{ a}_j}\left( {{ \sigma ^ + }{e^{i\left( {{\Delta _j} + q\nu } \right)t}} + { \sigma ^ - }{e^{i\left( {{\Delta _j} + q\nu - 2{\omega _{21}}} \right)t}}} \right){{ b}^\dag }^q{f_q}\left( {{{ n}_m},{\alpha _j}} \right)} \nonumber \\ 
&&\qquad + {h_j}{{ a}_j}\left( {{ \sigma ^ + }{e^{i{\Delta _j}t}} + { \sigma ^ - }{e^{i({\Delta _j} - 2{\omega _{21}})t}}} \right){f_0}\left( {{{ n}_m},{\alpha _j}} \right) \nonumber \\ 
&&\qquad+ H.c.,
\label{InteractionHamiltonian2} 
\end{eqnarray} 
\end{widetext} 
where ${\Delta _j} = \frac{{2g_j^2}}{\nu }\left( {{{ n}_j} + 1} \right) + {\omega _{21}} - {\omega _j}$ is an intensity-dependent detuning. As can be seen from the Hamiltonian (\ref{InteractionHamiltonian2}), with properly chosen ${\Delta _j}$, the rotating wave approximation (RWA) can be used to neglect the rapidly oscillating terms in the Hamiltonian (\ref{InteractionHamiltonian2}). If ${\Delta _j} $ is chosen to be $\pm q \nu$ (with $q$ as an integer number), the contributions associated with the second terms in each line of the Hamiltonian (\ref{InteractionHamiltonian2}) can be neglected using the RWA. Carrier resonance, red sidebands and blue sidebands occur for $q = 0$, $q>0$ and $q<0$, respectively . This simplified form of the Hamiltonian is very useful to describe single 
atom-assisted OMSs. The schematic energy-level spectrum of the hybrid OMS under consideration is presented in Fig. (\ref{fig:FIG1}-b).  
 
The approximate Hamiltonians corresponding to the carrier resonance, $\tilde {\cal H}_2^c$, the first red sideband, $\tilde {\cal H}_2^{r}$, and the first blue sideband, $\tilde {\cal H}_2^{b} $, are obtained by setting ${\Delta _j} = 0$, ${\Delta _j} = \nu$ and ${\Delta _j} =- \nu$ respectively, and they have the forms 
\begin{subequations} 
\label{eq:whole} 
\begin{equation} 
\tilde {\cal H}_2^c = {h_j}{f_0}\left( {{{ n}_m},{\alpha _j}} \right)\left( {{{ a}_j}{\sigma ^ + } +  a_j^\dag {\sigma ^ - }} \right), 
\label{carrierHamiltonian} 
\end{equation} 
\begin{equation} 
\tilde {\cal H}_2^{r} = -{h _j}{ { {\alpha _j}} }\left[ {{f_1}\left( {{{ n}_m},{\alpha _j}} \right){{ a}_j}{\sigma ^ + }{{ b}} +  a_j^\dag {\sigma ^ - }{{ b}^{\dag }}{f_1}\left( {{{ n}_m},{\alpha _j}} \right)} \right], 
\label{redHamiltonian} 
\end{equation} 
\begin{equation} 
\tilde {\cal H}_2^{b} = {h _j}\alpha _j\left[ {{\sigma ^ + }{{ a}_j}{{ b}^{\dag }}{f_1}\left( {{{ n}_m},{\alpha _j}} \right) + {f_1}\left( {{{ n}_m},{\alpha _j}} \right){\sigma ^ - } a_j^\dag {{ b}}} \right]. 
\label{blueHamiltonian} 
\end{equation} 
\end{subequations} 
It should be noted that the functions ${f_0}\left( {{{n}_m},{\alpha _0}} \right)$ and ${f_1}\left( {{{n}_m},{\alpha _j}} \right) $ can be written in terms of the associated Laguerre polynomials as follows 
\begin{subequations}
\begin{equation}
{f_0}\left( {{{n}_m},{\alpha _0}} \right) = {e^{ - \alpha _0^2/2}}\sum\limits_{n = 0}^\infty  {{L_n}\left( {\alpha _0^2} \right)\left| n \right\rangle \left\langle n \right|},
\end{equation}
\begin{equation}
{f_1}\left( {{{n}_m},{\alpha _j}} \right) = {e^{ - \alpha _j^2/2}}\sum\limits_{n = 0}^\infty  {L_n^1\left( {\alpha _j^2} \right)\frac{{\left| n \right\rangle \left\langle n \right|}}{{n + 1}}}. 
\end{equation}
\end{subequations}
The total Hamiltonian of an OMS consisting of several optical modes, is a combination of Eqs. (\ref{carrierHamiltonian} - \ref{blueHamiltonian}), if the parameter ${\Delta _j}$ for each mode is properly chosen. 
\section{\label{sec:SecIII}Sphere-coherent states} 
In order to be self-contained, in this section, we briefly discuss the SCSs introduced in \cite{mahdifar2} together with some of their nonclassical properties. 
The bosonic annihilation and creation operators associated with the generalized one-dimensional harmonic oscillator algebra can be obtained by a simple nonlinear extension of the harmonic oscillator algebra which have the form 
\begin{equation}
B = b\mathfrak{f}(n),\qquad  B^\dag = \mathfrak{f}(n)b^\dag , \qquad n = b^\dag b 
\label{A}
\end{equation}
where the deformation function, $\mathfrak{f}(n)$, is a Hermitian function of the number operator, $n$. The generalized operators $B$ and $B^\dag$ satisfy the commutation relation which depends on the deformation function, $\mathfrak{f}(n)$,
\begin{subequations} 
\begin{equation}
[n, B] = - B,
\end{equation}
\begin{equation}
[n, B^\dag] =  B^\dag,
\end{equation}
\begin{equation}
[ B, B^\dag] = ( n+1) \mathfrak{f}^2( n+1) -  n \mathfrak{f}^2(n).
\end{equation}
\end{subequations} 
The corresponding nonlinear coherent state, $\vert \chi; \mathfrak{f} \rangle $, is defined as eigenvector  of the deformed annihilation operator, 
\begin{equation}
B \vert \chi; \mathfrak{f} \rangle = \chi \vert \chi; \mathfrak{f} \rangle. \label{def}
\end{equation}
For instance, a spacial type of  nonlinear coherent states may realized as stationary states of the center-of-mass motion of a trapped and bichromatically laser-driven ion far from the Lamb-Dicke regime by the following deformation function and complex eigenvalue, $\chi$,
\begin{equation}
\mathfrak{f}(k) = L_k^1({\eta ^2}){\left[ {(k + 1)L_k^0({\eta ^2})} \right]^{ - 1}}, \qquad \chi  = \frac{{i{\Omega _0}}}{{\eta {\Omega _1}}},
\end{equation}
where $ L_k^m$ is an associated Laguerre polynomial and $\eta$ is the  Lamb-Dicke parameter and $\Omega _0$ and $\Omega _1$ are Rabi frequencies corresponding to the ion-laser interaction \cite{de Matos}. 

On the other hand, the algebra corresponding to  two coupled one dimensional  harmonic oscillators  on the flat surface can be considered as a deformed one dimensional oscillator algebra, which corresponds to the mapping  of  a  two dimensional harmonics oscillator on a sphere. In \cite{mahdifar2} the relation  between the deformation function
of deformed one dimensional harmonic oscillator and the sphere curvature is given. The deformation function of this algebra on flat space is given by, 
\begin{equation}
\mathfrak{f}_f(n)=\sqrt{N+1-n},
\end{equation}
where $N+1$ is the dimensionality of the Fock space corresponding to the energy eigenvalue.
 The curved-space deformation function has the form
\begin{equation}
\mathfrak{f}_s(n)=\mathfrak{f}_f(n)g(\lambda,n),
\end{equation}
where
\begin{equation} 
g(\lambda ,n) =\sqrt{[ {\lambda \left( {N + 1 - n} \right) + \sqrt {1 + \dfrac{\lambda ^2}{4}} }]({\lambda n + \sqrt {1 + \dfrac{\lambda ^2}{4}} })}. 
\end{equation} 
and $\lambda $ is the curvature of the sphere. 

The finite-dimensional coherent states associated with the sphere can be constructed as
\begin{equation} 
\left| \mu  \right\rangle  = {C^{1/2}}{e^{\mu {A^\dag }}}\left| 0 \right\rangle  = {C^{ - 1/2}}\sum\limits_{n = 0}^N {\sqrt {\left( {\begin{array}{*{20}{c}} 
N\\ 
n 
\end{array}}\right)} \left[ {g(\lambda ,n)} \right]!{\mu ^n}\left| n \right\rangle }, 
\end{equation} 
where $\mu$ is a complex number and $\left[ {g(\lambda ,n)} \right]! = g(\lambda ,n)g(\lambda ,n - 1) \cdots g(\lambda ,1)$ with $\left[ {g(\lambda ,0)} \right]! = 1$. The coefficient $C$ is obtained using the normalization condition as 
\begin{equation} 
C = \sum\limits_{n = 0}^N {\left( {\begin{array}{*{20}{c}} 
N\\ 
n 
\end{array}} \right){{\left\{ {\left[ {g(\lambda ,n)} \right]!} \right\}}^2}{\mu ^{2n}}}. 
\end{equation} 
\begin{figure*} 
\centering 
\includegraphics[width=5.5cm]{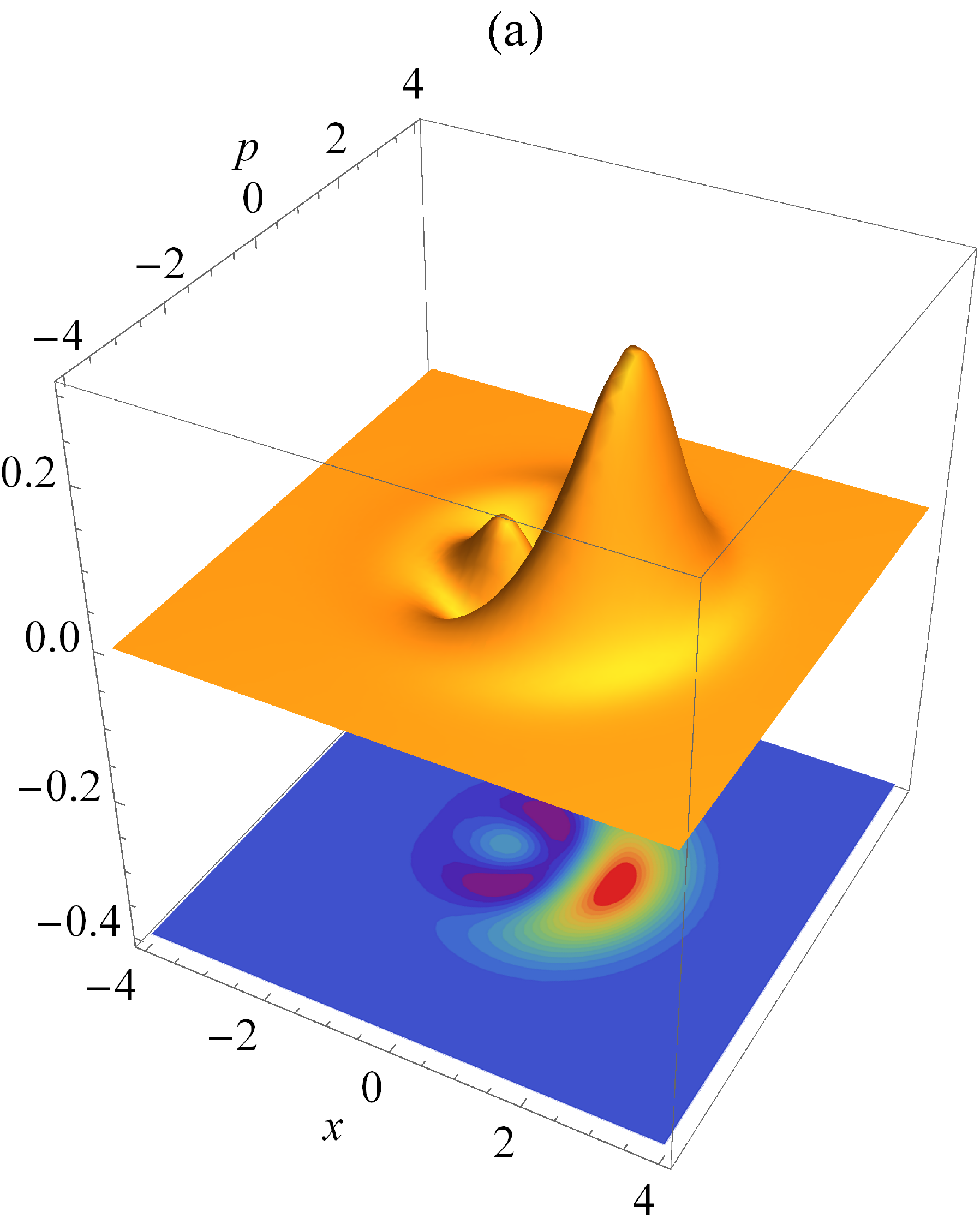} 
\includegraphics[width=5.5cm]{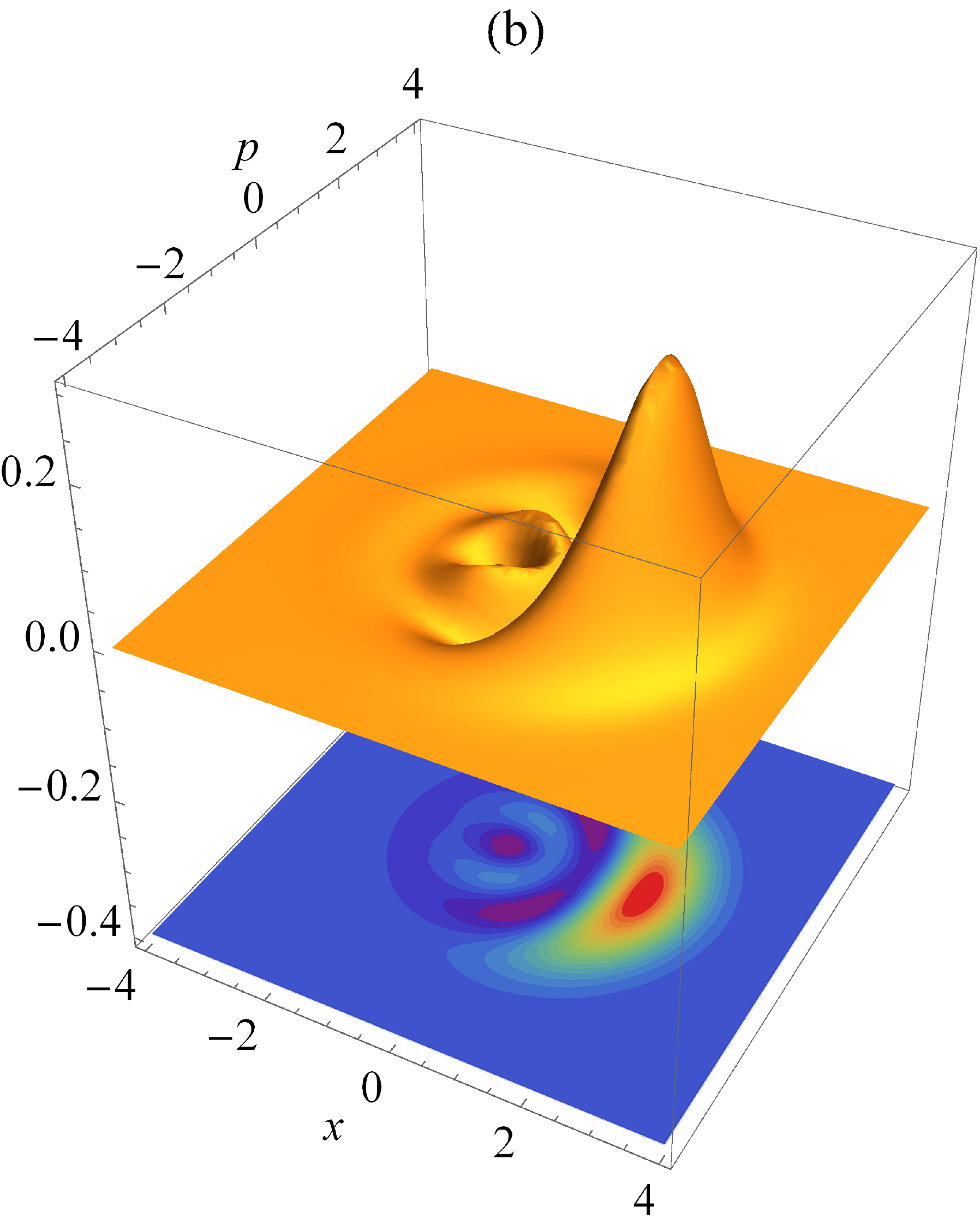} 
\includegraphics[width=5.5cm]{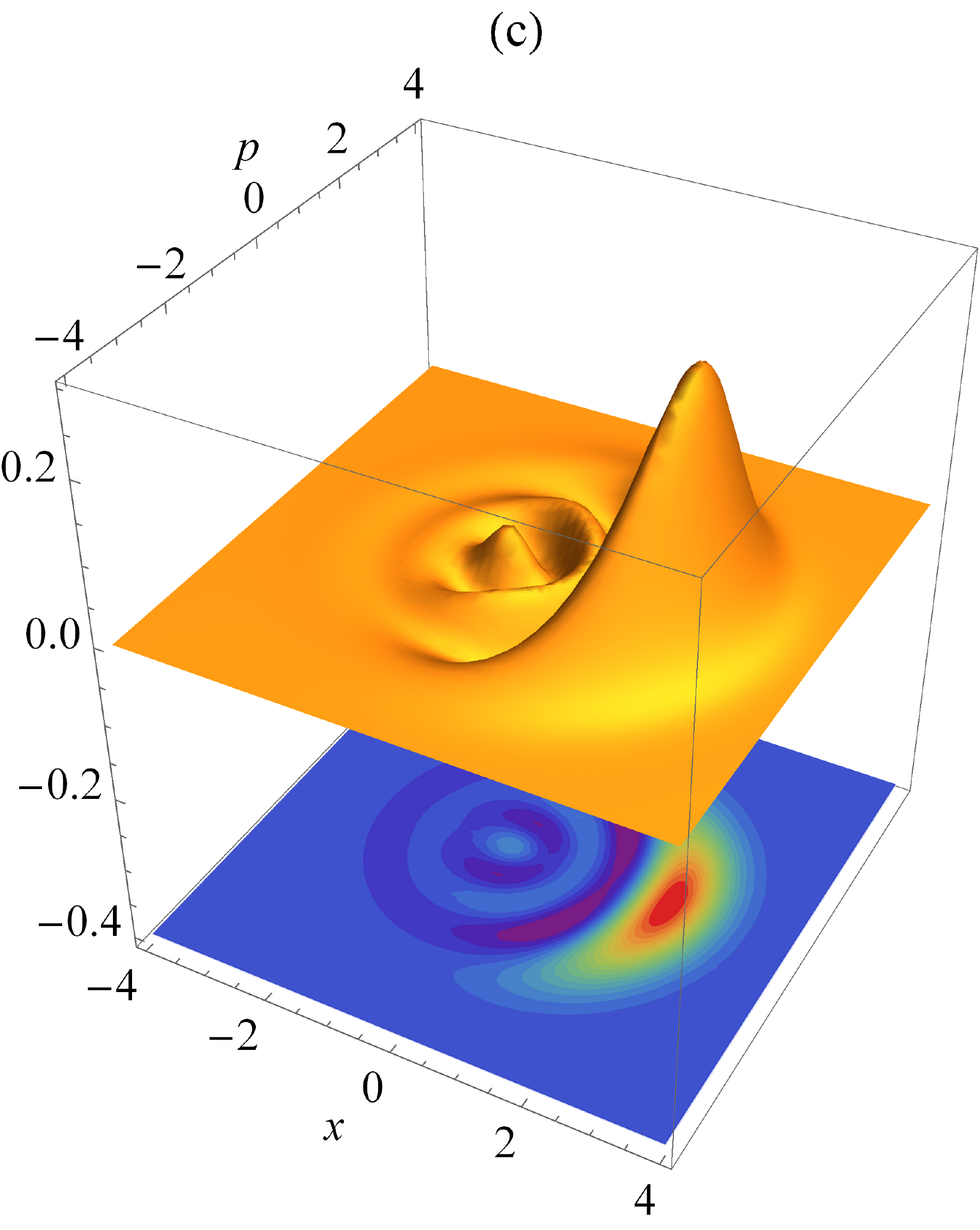} 
\caption{(Color online) Wigner distribution function of the SCSs with $\lambda=1$, $\mu=0.4$ and for different values of the Hilbert space dimension: (a) $N=2$, (b) $N=3$ and (c) $N=4$. As is seen, the Wigner 
function has negative values and shows interference structure over some region of the phase space.} 
\label{fig:FIG2} 
\end{figure*} 
\begin{figure} 
\centering 
\includegraphics[width=8cm]{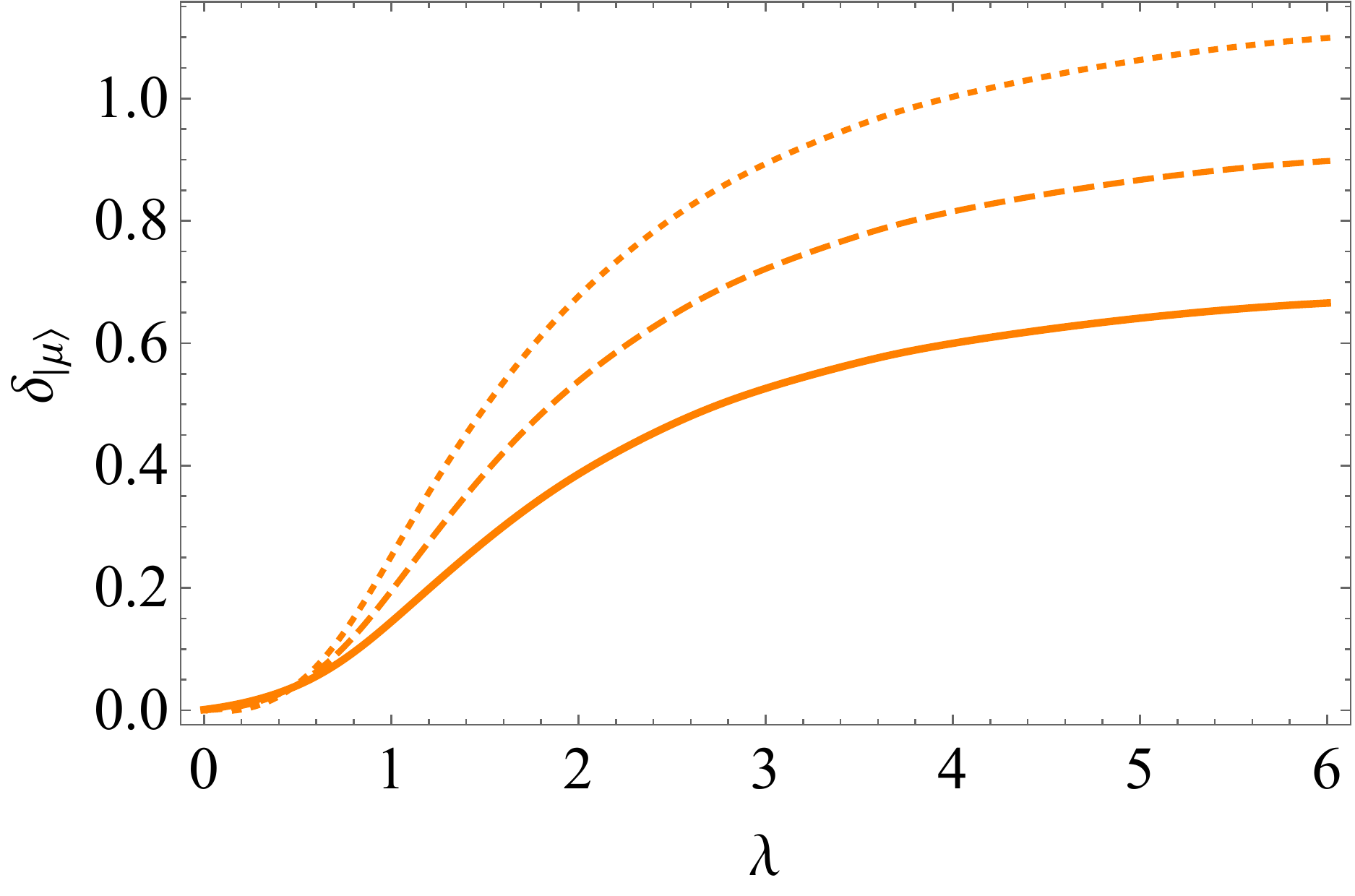} 
\caption{(Color online) The volume of the negative part of the Wigner function versus the sphere curvature for different values of the Hilbert space dimension: $N=2$ (solid line), $N=3$ (dashed line) and $N=4$ (dotted line).} 
\label{fig:FIG3} 
\end{figure} 

Some physical properties of the SCSs, in particular, their quantum statistical properties, including quadrature squeezing and the antibunching effect have been studied \cite{mahdifar,mahdifar2}. In this section the negativity of the Wigner distribution function and the degree of squeezing is studied as measures of the non-classicality of the SCSs. 

The Wigner distribution function for an SCS with density operator $\rho=\vert \mu \rangle \langle \mu \vert$ is given by
\begin{equation}
{{W_{\left| \mu \right\rangle }}\left( {x,p} \right)} \equiv \frac{1}{{2\pi }}\int_{ - \infty }^\infty  {\left\langle {{x + y/2}}
 \mathrel{\left | {\vphantom {{x + y/2} \mu }}
 \right. \kern-\nulldelimiterspace}
 {\mu } \right\rangle \left\langle {\mu }
 \mathrel{\left | {\vphantom {\mu  {x - y/2}}}
 \right. \kern-\nulldelimiterspace}
 {{x - y/2}} \right\rangle {e^{ipy}}dy}, 
\end{equation}
where $\vert x\pm y/2\rangle$ are the eigenvector of the position operator. 
The Wigner function of the SCSs of dimensions $N=2$, $N=3$, and $N=4$ is depicted in Fig (\ref{fig:FIG2}). As can be seen, the Wigner function becomes negative and possesses oscillatory behavior over some region of the phase space, which is a signature of non-classicality of the SCSs. The total number of peaks and holes is equal to $N+1$, i.e., the Wigner function exhibits more oscillations for higher Hilbert space dimensions. To quantify non-classicality of the SCSs, the volume of the negative part of the Wigner function is used as a measure \cite{negwign}. It is defined as 
\begin{equation} 
{\delta _{\left| \mu \right\rangle }} = \int {\int {\left| {{W_{\left| \mu \right\rangle }}\left( {x,p} \right)} \right|dxdp} } - 1. 
\end{equation}  
By definition, the quantity ${\delta _{\left| \mu \right\rangle }}$ is equal to zero for those states whose Wigner functions are non-negative (e.g., standard coherent states). 
In Fig. (\ref{fig:FIG3}), ${\delta _{\left| \mu \right\rangle }}$ is plotted as a function of $\lambda$ for different values of the Hilbert space dimension. One can see that with increasing the sphere curvature, $\lambda$, the nonclassicality of the state increases. Moreover, increasing the dimension of the Hilbert space leads to the enhancement of non-classicality of the states. 
 
\begin{figure} 
\centering 
\includegraphics[width=8cm]{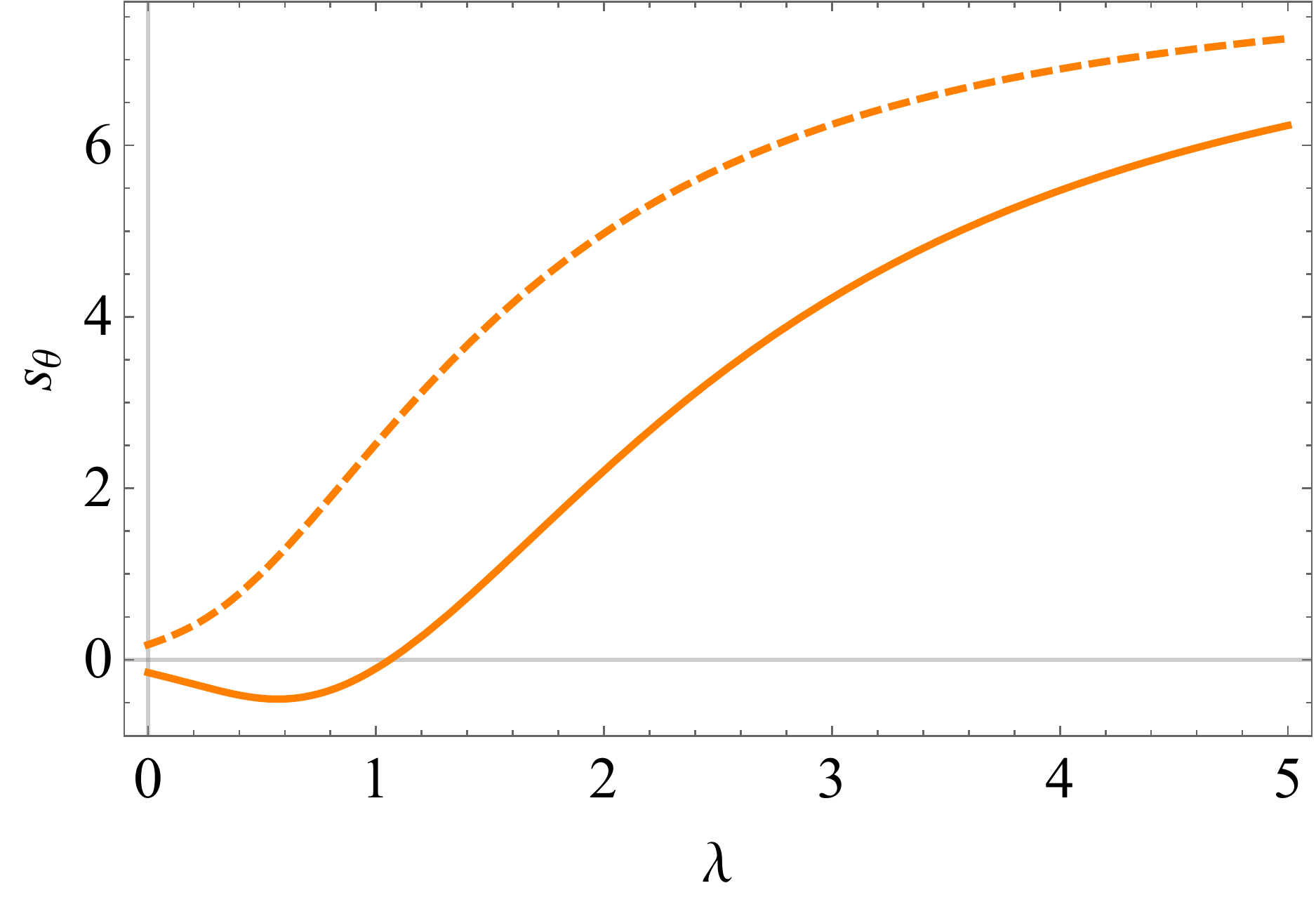} 
\caption{(Color online) The squeezing parameter as a function of the sphere curvature for the Hilbert space dimension $N=4$ and $\mu=0.4$, for ${X}_{0}$ (solid line) and ${X}_{\pi/2}$ (dashed line).} 
\label{fig:FIG4} 
\end{figure}

To examine the quadrature squeezing of the SCSs, we consider the squeezing parameter, $s_\theta$, defined by 
\begin{equation} 
s_{\theta}= 4 \langle \Delta {X}_{\theta}^2 \rangle -1, 
\end{equation} 
where ${X}_\theta$ is the generalized quadrature operator 
\begin{equation} 
{X}_{\theta}= \frac{1}{2} \left( {b} e^{-i \theta}+{b}^\dag e^{i \theta} \right). 
\end{equation} 
Quadrature squeezing occurs whenever the condition $-1 \leq s_{\theta} < 0$ is satisfied. As shown in Fig. (\ref{fig:FIG4}), SCSs exhibit quadrature squeezing for some values of $\lambda$. Reduced fluctuations in ${X}_{0}$ for relatively small values of sphere curvature is evident. A similar behavior for the SCSs with different Hilbert space dimension is also noticed. However, in the other quadrature, ${X}_{\pi/2}$, there is not such a reduction in fluctuations. Nonlinear interactions, for instance, parametric down conversion, are responsible for the occurrence of quadrature squeezing phenomenon in physical systems. In the present scheme, however, the nonlinear nature of the optomechanical coupling leads to such a noise reduction. 
\section{\label{sec:SecIV}Sphere-coherent motional states for the mechanical oscillator} 
In this section, we consider the realization of the sphere-coherent motional states for the MO in an atom-assisted optomechanical system. 
To illustrate the idea, we consider an MO interacting with $N+1$ optical modes of a Fabry-Perot cavity. 
In the case of the cavity field with $N+1$ independent modes, each mode may be described by Eqs. (\ref{carrierHamiltonian}-\ref{blueHamiltonian}) depending on its detuning. 
Detuning of each mode can be set in such a way that one of the modes (mode with index $j=0$) induces Hamiltonian (\ref{carrierHamiltonian}) and the others (modes with $j=1,2...,N$) induce Hamiltonian (\ref{redHamiltonian}). Consequently, the Hamiltonian of the total system in the interaction picture takes the form 
\begin{eqnarray} 
&& \tilde {\cal H} = \left[ { - \sum\limits_{j = 1}^N {{h _j}{\alpha _j}{f_1}\left( {{{ n}_m},{\alpha _j}} \right) b}{ a}_j + {h_0}{f_0}\left( {{{ n}_m},{\alpha _0}} \right) { a}_0} \right] \sigma ^{+} \nonumber \\ 
&& \qquad + H.c. \label{TotalHamiltonian} 
\end{eqnarray} 
In order to make the model under considearation more realistic, we should take into account the dissipation mechanisms involved in the system. Dissipative processes in the system can be studied using the master equation. Dissipation arises from the thermal heating of the MO vibration, the spontaneous emission of the atom and the photon leakage to the outside of the cavity. The master equation describing the system is given by ($\hbar=1$) 
\begin{equation} 
{\partial _t}\rho = - i\left[ \tilde {\cal H},\rho  \right] +{\cal L}_a \rho +{\cal L}_m \rho +{\cal L}_c \rho,
\end{equation} 
with 
\begin{subequations} 
\begin{equation}
{\cal L}_a \rho = \frac{{{\gamma _a}}}{2}{\cal L}[{ \sigma ^ - }]\rho ,
\end{equation}
\begin{equation}
{\cal L}_m \rho = \frac{{{\gamma _m}}}{2}\left( {{{\bar n}_m} + 1} \right){\cal L}[ b]\rho +\frac{\gamma _m}{2}{\bar n}_m {\cal L}[b]\rho ,
\end{equation}
\begin{equation}
{\cal L}_c \rho = \sum\limits_{j= 0}^N {\frac{{{\gamma _j}}}{2}{\cal L}[a_j]\rho }. 
\end{equation}
\label{comm}
\end{subequations} 
where $\gamma_m$, $\gamma _a$, and $\gamma _j$ are, respectively, the decay rates of the mechanical oscillator, the atom, and the optical modes and $\bar n_m={\left( {{e^{\nu /{k_B}T}} - 1} \right)^{ - 1}}$ denotes the mean number of phonons at temperature $T$.
 The Lindblad superoperator has the form 
\begin{equation} 
\mathcal{L}[{O}] \rho \equiv 2{O} \rho { O^\dag } - {{ O^\dag } }{O} \rho -  \rho { { O^\dag } }{O}. 
\end{equation} 
Typically, the spontaneous emission rate is much larger than the decay rates of  the MO and the optical modes. On the other hand, for cavities with high-Q factor the rate of photon leakage will be very small. Therefore, for times (comparable to the evolution of atomic variables) short enough that the photon leakage and thermal heating of the MO can be ignored ($\gamma _a\gg\gamma_m\gg\gamma_j $), only the spontaneous emission of the atom is taken into account. Using this assumption the master equation takes the form 
\begin{equation} 
{\partial _t}\rho = - i\left[ \tilde {\cal H},\rho  \right] +{\cal L}_a \rho.
\label{densityMatrix} 
\end{equation} 
For time intervals larger than the characteristic time of the atomic spontaneous emission and much smaller than the decay times of both the optical and the mechanical modes, one can assume that the atom reaches the steady state, i.e., its ground state, faster than the MO and the optical field. In the following, we first consider the case in which the decoherence of both the cavity and the mechanical modes can be neglected. In this way, we study the generation of the motional SCSs in the presence of the atomic decay as the main source of decoherence. Then, we examine the influence of the damping of the motion of the MO on the generated motional SCSs.  
\begin{figure} 
\centering 
\includegraphics[width=8cm]{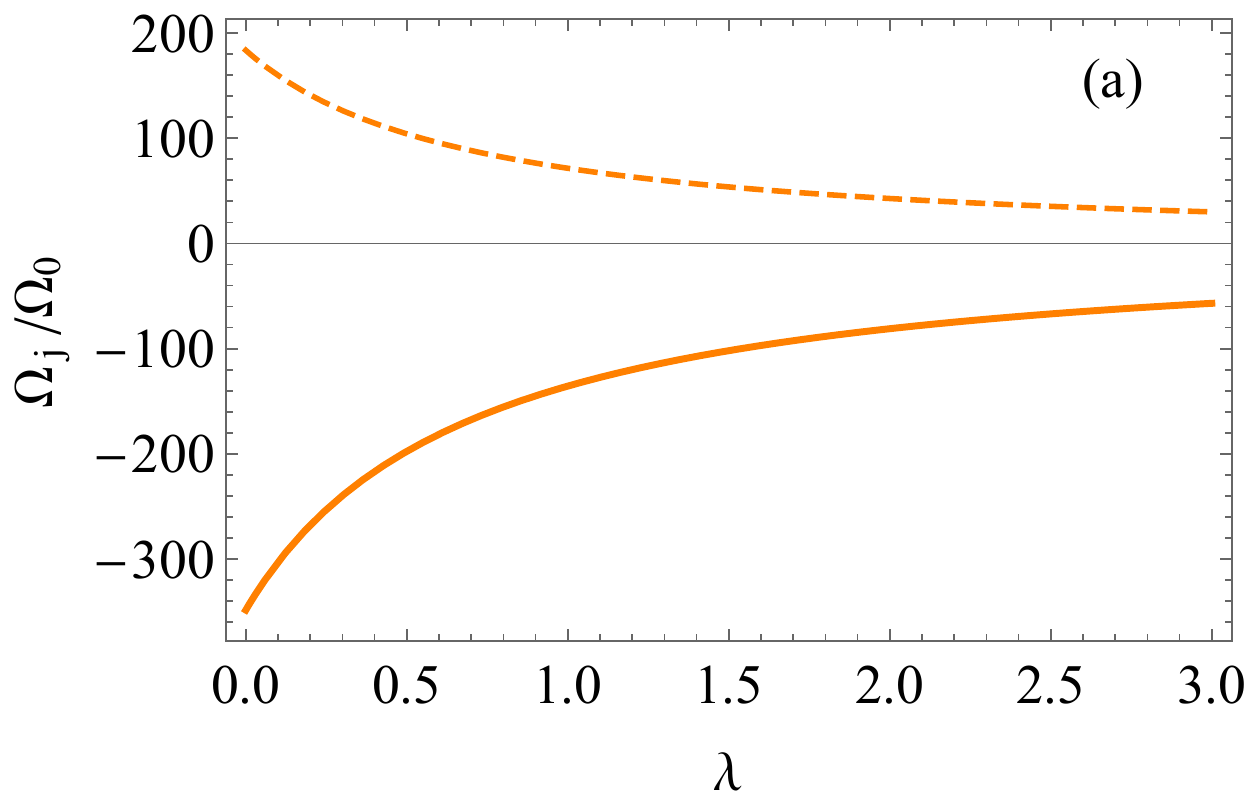} 
\includegraphics[width=8cm]{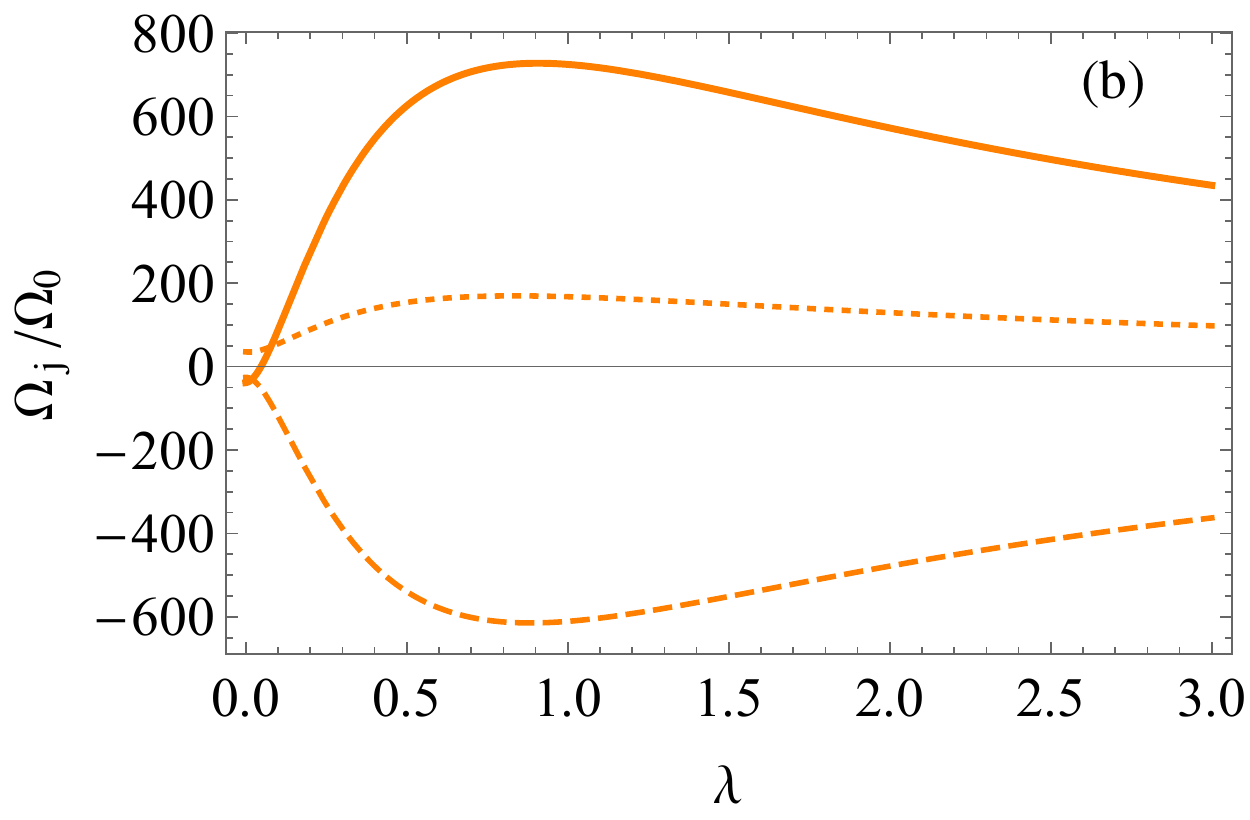} 
\includegraphics[width=8cm]{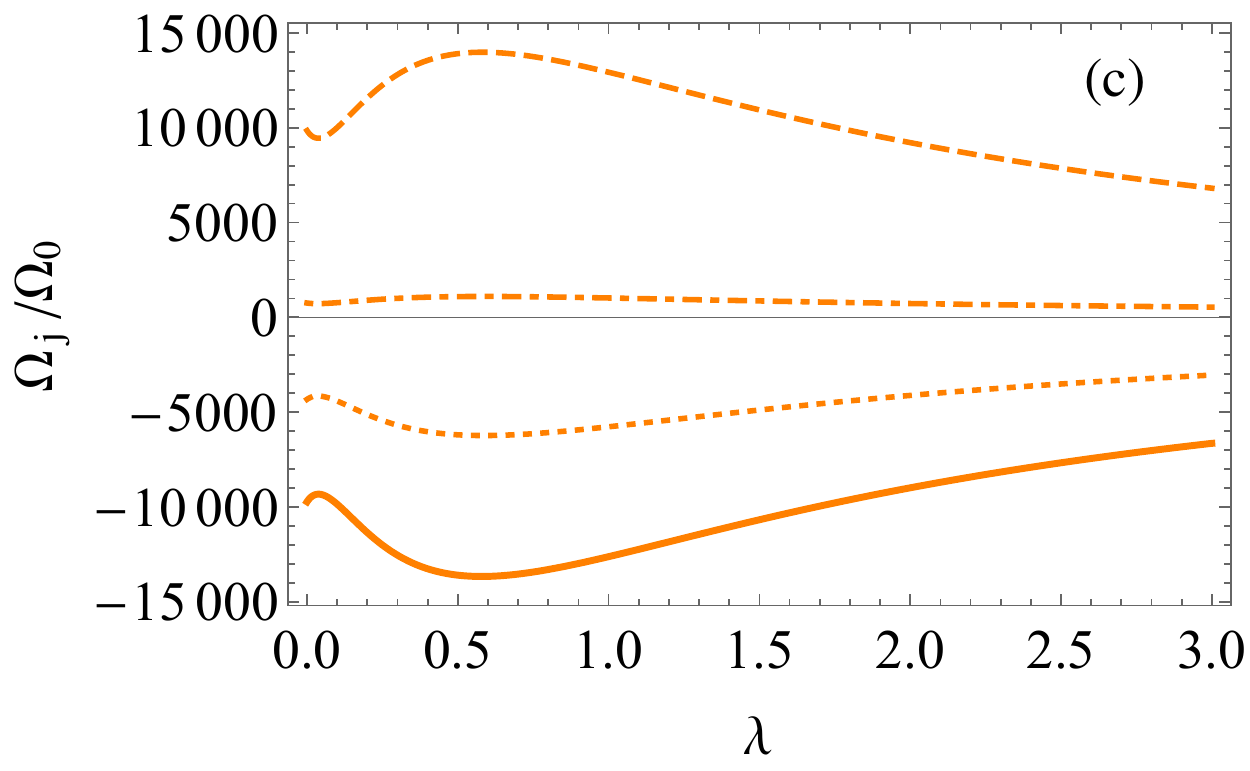} 
\caption{(Color online) The normalized Rabi frequency $\Omega_j/\Omega_0$ as a function of the curvature $\lambda$, for $\mu=0.4$ and different values of $N$: (a) $N=2$, (b) $N=3$, and (c) $N=4$. $\alpha _0$ is chosen so that, $L_{N}\left( {\alpha _0^2} \right)=0$, and the other optomechanical couplings are taken to be $j/10\quad (j=1,...,n)$. The solid, dashed, dotted and dashed-dotted lines correspond to $\Omega_1/\Omega_0$, $\Omega_2/\Omega_0$, $\Omega_3/\Omega_0$ and $\Omega_4/\Omega_0$, respectively.} 
\label{fig:FIG5} 
\end{figure} 
If the system is initially prepared in a pure state, after a short period of time the atomic subsystem reaches its steady state. The other two subsystems will be found in a pure entangled state because they are not still affected by dissipations from the environment. Therefore, the state of the system, after reaching the steady state of the atom, can be written as 
\begin{equation} 
{\rho _s} = \left| g \right\rangle \left| {{\psi _s}} \right\rangle \left\langle {{\psi _s}} \right|\left\langle g \right|, 
\end{equation}  
where $ \left| g \right\rangle$ is the atomic ground state and $\left| {{\psi _s}} \right\rangle$ is the pure entangled state of the MO-cavity field subsystem.  
Therefore Eq. (\ref{densityMatrix}) reduces to 
\begin{equation} 
\left[ { - \sum\limits_{j = 1}^N {{h_j}{\alpha _j}{f_1}\left( {{{ n}_m},{\alpha _j}} \right) b} {{ a}_j} + {h_0}{f_0}\left( {{{ n}_m},{\alpha _0}} \right){{ a}_0}} \right]\left| {{\psi _s}} \right\rangle = 0. 
\label{eq:eqo} 
\end{equation}
The entangled state $\left| {{\psi _s}} \right\rangle$ can be expressed in terms of continuous superposition of coherent states of the radiation field and discrete superposition of the MO Fock states: 
\begin{equation} 
\left| {{\psi _s}} \right\rangle = \int {{d^2}\{ {\beta _j}\} \left[ {\sum\limits_m {{C_m}(\{ {\beta _j}\} )\left| m \right\rangle \left| {\{ {\beta _j}\} } \right\rangle } } \right]}, 
\label{eq:superposition} 
\end{equation} 
where ${\left| {\{ {\beta _j}\} } \right\rangle }$ is the multimode coherent state. 
Inserting Eq.(\ref{eq:superposition}) into Eq. (\ref{eq:eqo}) the following recurrence relation is obtained 
\begin{equation} 
{C_n}(\{ {\beta _j}\} ) = \sqrt {n !} \prod\limits_{i = 0}^{n - 1} {\frac{{{\Omega _0}\left( {{e^{ - \alpha _0^2/2}}{L_i}\left( {\alpha _0^2} \right)} \right)}}{{\sum\limits_{j = 1}^N {{\Omega _j}{\alpha _j}{e^{ - \alpha _j^2/2}}L_{i}^{(1)}\left( {\alpha _j^2} \right)} }}} {C_0}(\{ {\beta _j}\} ), 
\end{equation} 
where $\Omega_j=h_j \beta_j$ is the Rabi frequency corresponding to the atom-field coupling. If the Lamb-Dicke parameter, $\alpha_0$, is chosen such that $L_N(\alpha_0^2)=0$, then the dimension of the Hilbert space is bounded from the upper limit and a Hilbert space with dimension of $N+1$ is obtained. Using a similar procedure as in \cite{mahdifar}, we arrive after some calculations  at the following set of algebraic equation for given $\alpha_j (j=1,...,n)$ and $N$ unknown quantities $\xi_j$
\begin{eqnarray} 
&&\sum\limits_{j = 1}^N {{\Omega _j}{\alpha _j}{e^{ - \alpha _j^2/2}}L_{n - 1}^1\left( {\alpha _j^2} \right)} {\xi _j} = \quad \quad \quad \quad \quad \quad \quad \quad \quad \quad \quad \quad \nonumber\\ 
&&\quad \quad \quad \quad \quad \quad \quad \quad \frac{n}{{\sqrt {\left( {N - n + 1} \right)} }}\frac{1}{{g(\lambda ,n)}}{L_{n - 1}}\left( {\alpha _0^2} \right), 
\label{eq:eql} 
\end{eqnarray} 
where 
\begin{equation} 
{\xi _j} = \mu {e^{\alpha _0^2/2}}\left( {{\Omega _j}/{\Omega _0}} \right). 
\end{equation} 
For given optomechanical couplings $\alpha_j (j=1,...,n)$ Eq. (\ref{eq:eql}) can be solved. Each set of the atom-field and the optomechanical couplings which satisfies Eq. (\ref{eq:eql}), will produce an SCS with specified curvature and amplitude. Therefore, using (\ref{eq:eql}) allows us to obtain unknown quantities $\xi_j (j=1,...,n)$. 

The solution of Eq. (\ref{eq:eql}) for the Hilbert space dimensions of $N=2$, $N=3$, and $N=4$ as a function of the curvature is presented in Fig. (\ref{fig:FIG5}). It should be noted that the negative values of the ${{\Omega _j}/{\Omega _0}}$ correspond to a $\pi$ phase shift of the optical mode. 
Therefore, controlling the optomechanical and the atom-field interactions enables us to generate motional SCSs for the MO. 

Regarding the deformation function $\mathfrak{f}\left( {{n_m}} \right)$ associated with the motional SCSs obtained from Eq. (\ref{eq:eql}), it is easy to show that
\begin{equation}
\mathfrak{f}\left( {{n_m}} \right) = \frac{{\sum\limits_{j = 1}^N {{\beta _j}{h_j}{\alpha _j}{f_1}\left( {{n_m},{\alpha _j}} \right)} }}{{{\beta _0}{h_0}{f_0}\left( {{n_m},{\alpha _0}} \right)}},
\end{equation}
Where $\beta_j$s denote the amplitude of the multi-mode coherent states of the cavity field.
\section{\label{sec:SecV} Effect of the mechanical oscillator damping on the generated motional SCS$\rm s$} 
\begin{figure} 
\centering 
\includegraphics[width=8cm]{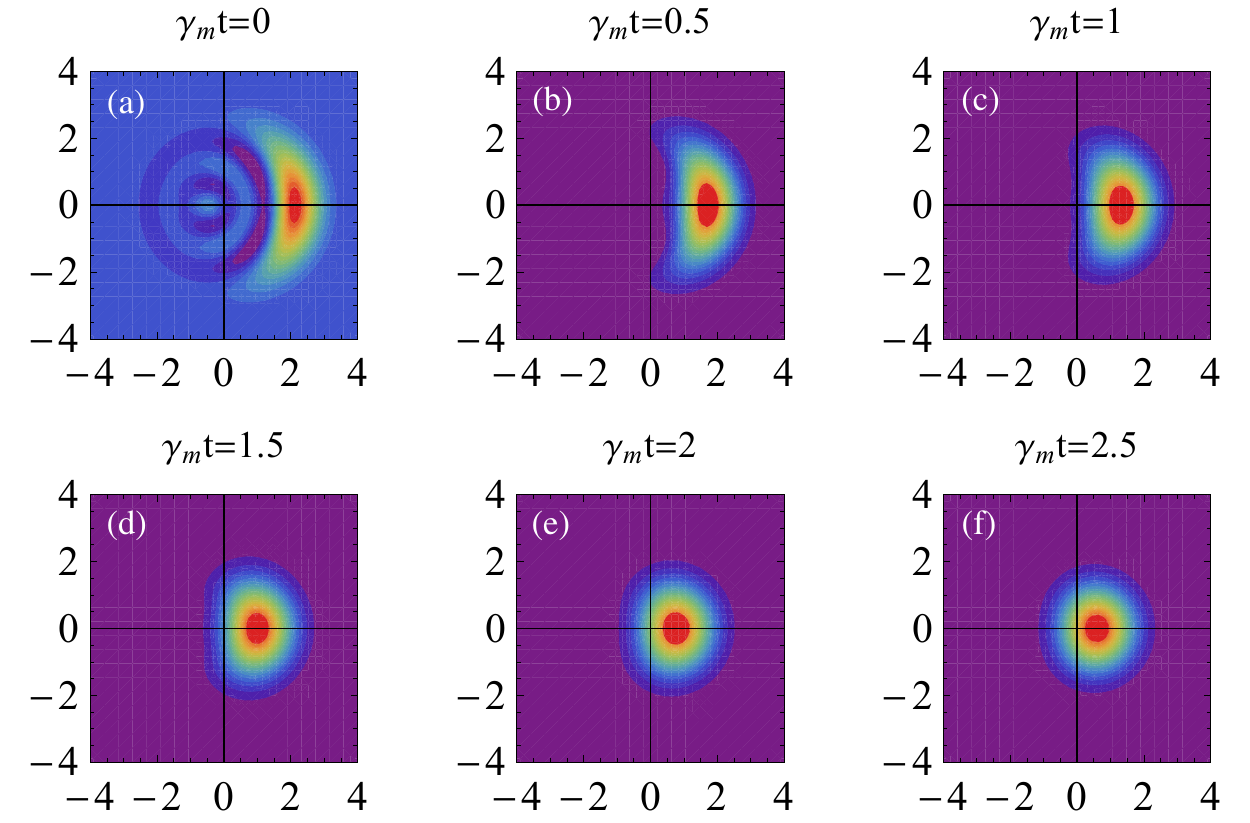} 
\includegraphics[width=8cm]{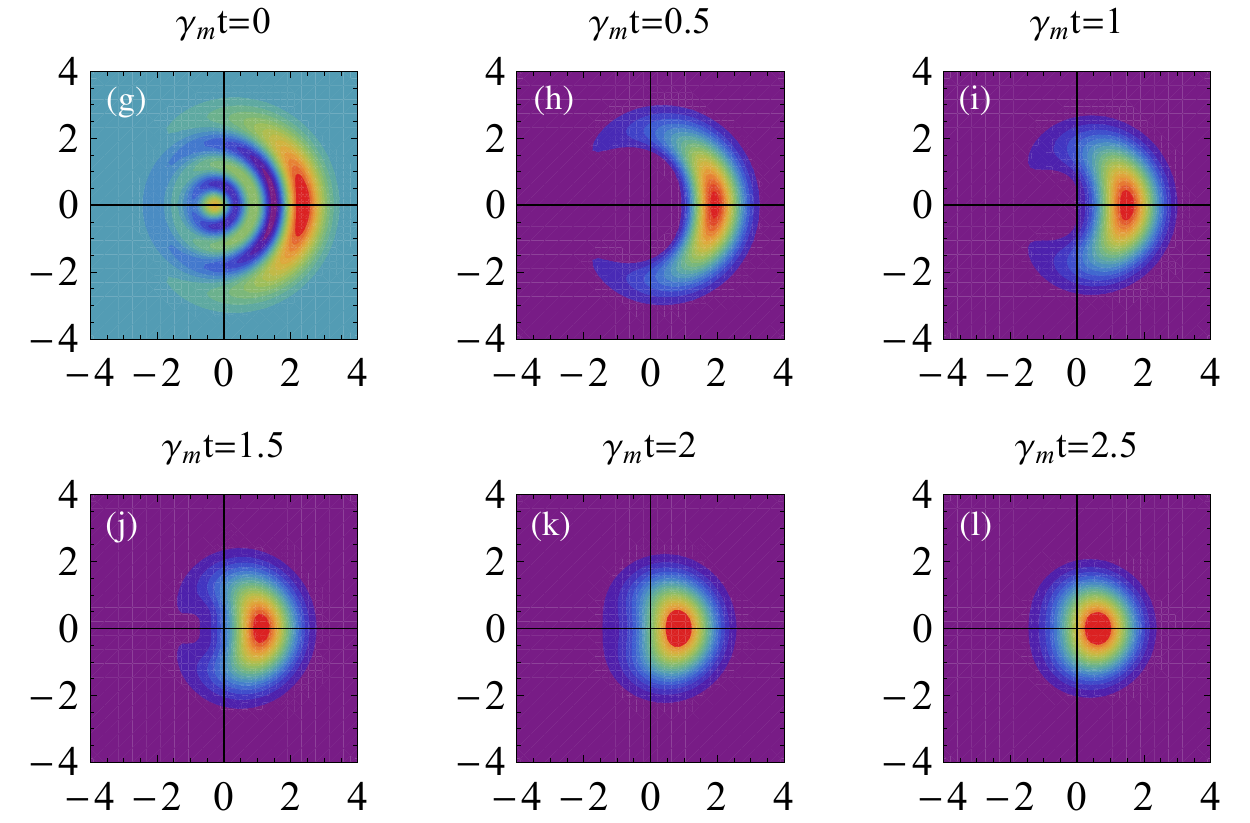} 
\caption{(Color online) Evolution of the Wigner function of the motional SCSs with $\mu=0.4$, $\bar n_m=0.5$, and for $\lambda=1$ (a-e) and $\lambda=2$ (g-l). Here we have set $N=4$. } 
\label{fig:FIG6} 
\end{figure}

Separating time scales of dissipations enables us to investigate the damping mechanisms involved in the problem. As mentioned in Sec. \ref{sec:SecIV}, the SCSs can be created for time scales much larger than the characteristic time of atomic decay, which means that the damping of the atom is desirable in this case. In other words, the quantum coherence is achieved by a decoherence process. However, the damping of the MO due to its coupling to the thermal heat bath destroys the induced coherence in the system. To gain more insight into the effect of the damping of the mechanical motion on the mirror SCSs, we transform the reduced master equation for the density operator of the mirror $\rho_m$, i.e., 
\begin{equation} 
{\partial _t}{\rho _m} = \frac{{{\gamma _m}}}{2}\left( {{{\bar n}_m} + 1} \right){\cal L} [ b]\rho_m + \frac{{{\gamma _m}}}{2}{\bar n_m}{\cal L}[b]\rho_m {\rm{ }}. 
\end{equation} 
into the Fokker-Planck equation to examine the time evolution of the generated motional SCSs under different initial conditions. The Fokker-Planck equation for the damped mechanical oscillator in the Wigner representation reads \cite{Carmichael}  
\begin{eqnarray}
&&\frac{\partial }{{\partial t}} W(x,p) = \frac{{{\gamma _m}}}{2}\left( {\frac{\partial }{{\partial x}}x + \frac{\partial }{{\partial p}}p} \right)W(x,p) \qquad \qquad \qquad \quad\nonumber\\
&&\qquad \qquad \quad+ \frac{{{\gamma _m}}}{4}\left( {{{\bar n}_m} + \frac{1}{2}} \right)\left( {\frac{{{\partial ^2}}}{{\partial {x^2}}} + \frac{{{\partial ^2}}}{{\partial {p^2}}}} \right)W(x,p).
\end{eqnarray}
The time evolution of the Wigner function for a motional SCS with $\mu= 0.4$ for $\lambda= 1$ and $\lambda= 2$, and for $N=3$ is shown in Fig. (\ref{fig:FIG6}). As the system evolve, the fringes corresponding to the superposition of the Fock states disappear, finally, Wigner function of a thermal state with ${\bar n}_{m}=0.5$ is reached. As can be seen, the state with greater $\lambda$ is more robust against the damping mechanism because under the same condition, the nonclassicality of the state with $\lambda= 2$ is larger than that of a state with $\lambda= 1$. To obtain a state more robust against dissipation, one could increase the Hilbert space dimension or curvature of the space (see Fig. (\ref{fig:FIG3})). 

It should be noted that the MO dissipation mechanism destroys the quantum coherence in the SCSs. But the quantum coherence associated with SCSs is created by another dissipative channel in the system, i.e., the atomic spontanoues decay. The generated coherence in the motional state of the MO may be regarded as a competition between the unitary and nonunitary evolution terms involved in Eq. (\ref{densityMatrix}). Therefore, in the proposed scheme, the dissipation mechanisms play a dual role. 

\begin{figure} 
\includegraphics[width=8cm]{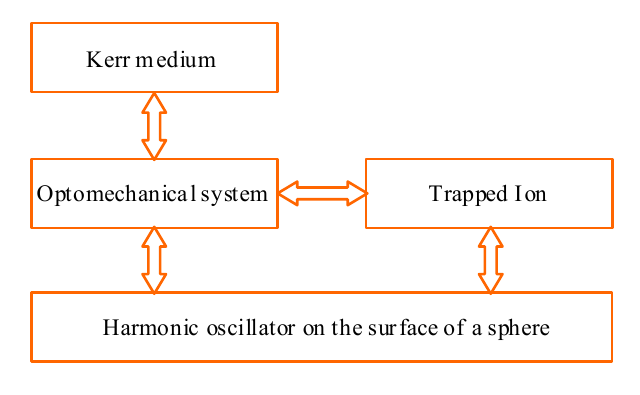} 
\caption{(Color online) Schematic diagram of the analogies that have been used in this paper.} 
\label{fig:FIG7} 
\end{figure} 
\section{\label{sec:SecVI}Conclusions} 
The main idea of our proposal can be summarized as
follows. In an atom-assisted optomechanical cavity, it is possible to control the interaction in order to obtain an effective Hamiltonian given by Eqs. (\ref{carrierHamiltonian} - \ref{blueHamiltonian}). In our treatment the intrinsic
non-linearity of the optomechanical interaction plays an
essential role. This provides a flexible platform for generation and investigation of non-classical motional states of the mechanical element. A specific class of nonlinear coherent states, the so-called sphere coherent motional state, can be achieved by controlling the atom-field and MO-field interactions in a multi-mode optomechanical cavity. These states are non-Gaussian and possess negative Wigner function. They also exhibit quadrature squeezing. 

As is shown in Fig. (\ref{fig:FIG7}), three analogies have been used in this paper. The first one is the analogy between a Kerr medium and an OMS which allows us to introduce an intensity-dependent detuning. The second analogy is the analogy between an OMS and a trapped ion which can be understood from their Hamiltonians. This analogy is significant and leads to cross-fertilization of ideas in OMSs and trapped ion systems. Using this analogy the atom-assisted OMSs can be applied in the same manner as the trapped ion system, e.g., quantum information processing or quantum simulation.  The third one is the analogy between a harmonic oscillator on the surface of a sphere and the OMS as well as the trapped ion systems. In addition, the dissipation mechanisms involved in the problem lead to creation and destruction of the SCSs, that is, they play a dual role. 

The procedure introduced in this paper, provides a platform for quantum state engineering of a wide variety of nonlinear coherent states of the MO. For instance, in the case of two-mode cavity field, detuning of each mode can be adjusted so that the first mode is in resonance with the carrier and the second one resonantly drives the second red sideband. Working in the regime of the small Lamb-Dicke parameter (${f_q}\left( {{{\hat n}_m},{\alpha _j} \ll 1} \right) \simeq 1/q!$) the Hamiltonian describing the system takes the form 
\begin{equation}
{\cal H} = \frac{{{h_1}\alpha _1^2}}{2}\left( {{{\hat a}_1}{{\hat b}^2} + \frac{{2{h_0}}}{{{h_1}\alpha _1^2}}{{\hat a}_0}} \right){\sigma ^ + } + H.c.
\end{equation}
which is suitable for the  the generation of the motional Schr\"{o}dinger cat state \cite{VogelB}.

As an outlook, the present scheme could in principle be used for coherent state transfer between the MO and the motion of a single trapped atom. We hope to report on such issue in a forthcoming paper.

\end{document}